\documentclass[pdflatex,sn-mathphys-num]{sn-jnl}

\usepackage{graphicx}%
\usepackage{multirow}%
\usepackage{amsmath,amssymb,amsfonts}%
\usepackage{amsthm}%
\usepackage{mathrsfs}%
\usepackage[title]{appendix}%
\usepackage{xcolor}%
\usepackage{textcomp}%
\usepackage{manyfoot}%
\usepackage{booktabs}%
\usepackage{algorithm}%
\usepackage{algorithmicx}%
\usepackage{algpseudocode}%
\usepackage{listings}%
\usepackage{amsmath}
\usepackage{array}         
\usepackage{booktabs}       
\usepackage{multirow}      
\usepackage{enumitem}
\usepackage{subcaption}
\usepackage{dsfont}
\usepackage{multirow}

\theoremstyle{thmstyleone}%
\theoremstyle{thmstyletwo}%

\theoremstyle{thmstylethree}%

\begin{document}

\title[Article Title]{Routing-Aware Explanations for Mixture of Experts Graph Models in Malware Detection}

\author[1]{\fnm{Hossein} \sur{Shokouhinejad}}

\author[1]{\fnm{Roozbeh} \sur{Razavi-Far}}

\author[1]{\fnm{Griffin} \sur{Higgins}}

\author[1]{\fnm{Ali A} \sur{Ghorbani}}

\affil*[1]{\orgdiv{University of New Brunswick}, \orgname{Faculty
 of Computer Science}, \orgaddress{\street{3 Bailey Dr.}, \city{Fredericton}, \postcode{E3B5A3}, \state{New Brunswick}, \country{Canada}}}

\abstract{Mixture-of-Experts (MoE) offers flexible graph reasoning by combining multiple views of a graph through a learned router. We investigate routing-aware explanations for MoE graph models in malware detection using control flow graphs (CFGs). Our architecture builds diversity at two levels. At the node level, each layer computes multiple neighborhood statistics and fuses them with an MLP, guided by a degree reweighting factor $\rho$ and a pooling choice $\lambda \in \{\text{mean}, \text{std}, \text{max}\}$, producing distinct node representations that capture complementary structural cues in CFGs. At the readout level, six experts, each tied to a specific ($\rho$, $\lambda$) view, output graph-level logits that the router weights into a final prediction. Post-hoc explanations are generated with edge-level per expert and aggregated using the router’s gates so the rationale reflects both what each expert highlights and how strongly it is selected. Evaluated against single-expert GNN baselines such as GCN, GIN, and GAT on the same CFG dataset, the proposed MoE achieves strong detection accuracy while yielding stable, faithful attributions under sparsity-based perturbations. The results indicate that making the router explicit and combining multi-statistic node encoding with expert-level diversity can improve the transparency of MoE decisions for malware analysis.}


\maketitle

\section{Introduction}

The growing scale and increasing complexity of modern malware now require detection methods that are more advanced than basic signature-based systems. Modern malware detection systems generally rely on three families of techniques \citep{Our_Survey}. Signature and indicator matching are fast and precise, but vulnerable to packing and minor code changes. Static analysis inspects files without running them by extracting strings, opcodes, metadata, etc. Although this technique scales well, it struggles with obfuscation and code that activates itself only at runtime. Dynamic analysis executes a sample in a controlled environment in order to monitor what it actually does (e.g. processes it spawns, files it touches, network activity, and the control flow it traverses). This runtime observation is more difficult for adversaries to falsify consistently and captures behaviors that static methods may miss, such as unpacking, late binding, and environment-triggered logic. While dynamic approaches come with a higher operational cost, they provide far richer, behavior-focused evidence. As a result, dynamic malware detection has become an essential component in operational workflows, along with static and signature-based methods. Given this shift toward behavior-centric evidence, graph-based learning methods offer a natural way to model program structure observed at runtime.

Recently, machine learning has been at the core of many malware detection systems, shifting the focus from signatures and rules to models that learn patterns of behavior and structure \citep{survey}. Within this shift, graph neural networks (GNNs) have become more highlighted because program artifacts are naturally graph-structured. Graphs like control flow graphs (CFGs)~\citep{CFG_1}, function call graphs (FCGs)~\citep{FCG_1}, API call graphs~\citep{API_Call_Graph_1}, and system call interaction graphs~\citep{system_call} each capture a distinct facet of program execution. GNNs propagate information over nodes and edges, which allows a model to capture long-range dependencies, contextual effects, and compositional behavior that flat feature vectors struggle to represent. This is especially useful against techniques like packing, obfuscation, or code reuse, because the underlying relational structure of the code often stays intact and reveals the true intent. In practice, GNNs match well with standard security preprocessing: they can use per-node features (e.g., opcodes, API categories, basic-block statistics) and per-edge attributes (types of control transfer or call linkage), integrating them into robust graph embeddings. Therefore, GNN-based approaches have become a strong foundation for learning malware detectors that operate on rich program graphs and support downstream tasks such as classification~\citep{Consistency}. Yet the diversity of malware behaviors and graph structures motivates approaches that adapt capacity to each sample rather than relying on a single model.

Ensemble learning improves predictive performance by combining multiple models instead of relying on a single model~\citep{Stack}. Common families include bagging, which trains models on resampled data to reduce variance (for example, random forests); boosting, which adds models sequentially to correct residual errors and reduce bias (such as AdaBoost or gradient boosting); and stacking, which learns a meta-learner to fuse base model outputs. Mixture of Experts (MoE) extends ensembles with conditional computation: a set of experts is trained alongside a gating network that assigns input-dependent weights to experts, either densely with a softmax over all experts or sparsely with Top-k routing and load balancing (LB)~\citep{MoE_1}. This design encourages specialization, letting different experts handle different regions of the input space. In graph settings, MoE can be paired with GNNs so that experts specialize to distinct structural or behavioral patterns observed in program graphs, making the approach a natural fit for malware detection where samples vary widely across families, packers, and runtime behaviors~\citep{MoE_2,MoE_3}. As models route decisions through specialized experts, understanding why a prediction was made becomes essential.

Explainability makes machine learning outputs understandable by showing why a model reached a decision, not just what it predicted. It improves trust, supports auditing and error analysis, and helps analysts detect shortcuts or bias that raw accuracy can hide. In GNNs, explainability reveals which substructures, edges, or node attributes most influenced the prediction, turning a learned representation over a graph into human-interpretable evidence using techniques such as gradient and attribution methods (e.g., Integrated Gradients), perturbation or masking approaches that measure output change when structure is altered, and surrogate models that approximate local decision boundaries~\citep{Dual}. This is especially important when GNNs analyze program graphs, where explanations can point to concrete execution patterns rather than opaque feature vectors. In malware detection with GNNs, clear attributions help analysts validate alerts, prioritize investigations, and document findings with defensible rationale tied to program behavior. 

In this paper, we develop and study routing-aware, post-hoc explanations for MoE graph models applied to malware detection on CFGs. The approach begins with the dynamic symbolic extraction of CFGs from Portable Executable (PE) files. Each basic block is then embedded through a two-step feature representation. Our MoE promotes diversity at two levels: at the node level, each GNN layer computes multiple neighborhood statistics and fuses them with a lightweight MLP, controlled by a degree reweighting factor $\rho$ and a pooling choice $\lambda \in \{\mathrm{mean}, \mathrm{std}, \mathrm{max}\}$; at the readout level, six graph experts, each tied to a specific $(\rho,\lambda)$ view, produce graph-level logits that a gating network combines using Top-$k$ routing with LB. For explanations, we use a post-hoc procedure: edge-level explanations are computed per expert and then aggregated with the router’s gates so the rationale reflects both what each expert highlights and how strongly it contributes to the prediction. We evaluate against single-expert GNN baselines on shared splits and assess accuracy alongside routing-aware faithfulness under sparsity perturbations. The main contributions of this study are as follows:

\begin{itemize}
  \item An end-to-end framework for routing-aware explainable malware detection that unifies CFG construction, two-stage MoE--GNN modeling, Top-$k$ routing, and post-hoc expert-aligned attributions, together with a consistent evaluation setup for accuracy and routing-aware faithfulness.
  \item A two-level diversity MoE for CFG-based malware detection that combines multi-statistic node encoding $(\rho,\lambda)$ with six specialized experts and Top-$k$ routing.
  \item A routing-aware post-hoc explanation method that aggregates per-expert scores with the learned gates, aligning explanations with the model’s actual combination of experts.
\end{itemize}

The remainder of this paper is organized as follows. Section~\ref{sec:related_works} surveys prior research on GNN-based malware detection, ensemble methods with an emphasis on MoE, and explainability techniques for graph models. Section~\ref{sec:framework} details the proposed MoE framework, including CFG extraction from PE files, the two-step node feature embedding, the multi-statistic node encoding, graph-level readout with six experts, the Top-$k$ gating and LB strategy, and the routing-aware post-hoc explanation pipeline. Section~\ref{sec:result} describes the experimental setup, datasets, training configuration, and evaluation metrics, then presents the detection results and the explainability assessment. Finally, Section~\ref{sec:conclusion} concludes the paper and outlines directions for future work.

\section{Related Works}\label{sec:related_works}

Researchers are increasingly turning to graph-based methods to spot malware, especially by modeling a program's structure as a CFG. GNNs excel here, as they can decode the complex relationships within these graphs to identify malicious code. To make these systems even stronger and more reliable, some are combining multiple models using ensemble methods, particularly MoE frameworks. However, as the models grow more sophisticated, we also need to understand why they make their decisions. This "explainability" is crucial for real-world security applications. This section will cover the key developments in three areas: GNNs for malware detection, ensemble methods for analysis, and techniques for explaining what GNNs find.

Recent studies have increasingly exploited GNNs for malware detection, using CFGs as structural representations of executable behavior. Zhang et al.~\citep{CFG_1} designed a few-shot malware classification model employing a triplet-trained graph transformer that encodes each malware sample as a CFG. By optimizing a triplet loss function, their model learns embeddings that effectively capture both structural and semantic relationships among basic blocks, enabling better generalization in low-data settings. Peng et al.~\citep{MalGNE} further addressed the challenge of instruction-level representation by proposing MalGNE, a node embedding framework that first encodes assembly instructions using a rule-based vectorization method to mitigate out-of-vocabulary issues. The framework then applies aggregation and attention-based bidirectional LSTM layers to preserve the sequential and semantic dependencies of instructions within basic blocks before feeding them into a GNN for classification.

Building on these efforts, Amer et al.~\citep{new_2} developed GraphShield, a dynamic graph-based malware detection framework that models temporal API call interactions and employs attention-enhanced GNN architectures for behavioral analysis. GraphShield also integrates post-hoc explainability tools that identify critical API subgraphs responsible for malicious behavior, achieving higher accuracy and interpretability compared to conventional sequential models. Wang et al.~\citep{new_3} proposed SIGDroid, a scalable Android malware detection framework that uses knowledge distillation to transform heterogeneous FCGs into homogeneous representations while retaining semantic and permission-related information. Their student-centered distillation process allows compact GNNs to maintain strong performance across large-scale datasets, improving scalability and efficiency. In the context of critical infrastructure protection, Esmaeili et al.~\citep{new_1} introduced an adversarial IoT malware detection framework that leverages GNNs to identify manipulated CFGs prior to classification. Their approach successfully detects adversarially perturbed graphs from well-known attack families such as Gafgyt, Mirai, and Tsunami, attaining over 98\% detection accuracy and emphasizing the relevance of adversarial robustness in securing industrial and IoT systems.

MoE has recently been adapted to graphs to enable conditional computation and structural specialization. Wang et al.~\citep{MoE_1} introduce Graph Mixture of Experts (GMoE), where multiple aggregation experts are placed in each GNN layer and a gate selects experts per node, improving accuracy on OGB benchmarks while keeping inference efficient. Their study motivates expert specialization across different neighborhood ranges and shows consistent gains over single-expert baselines. In molecular learning, TopExpert~\citep{MoE_2} combines a standard GNN encoder with several lightweight expert heads; a clustering-based gating module routes each molecule to topology-specific experts, and the final prediction aggregates expert outputs with learned weights. For time-series anomaly detection, Huang et al.~\citep{MoE_3} propose a Graph-MoE that fuses hierarchical information from multiple GNN layers and augments the router with a memory unit that stores global historical patterns to weight experts adaptively; extensive experiments across five benchmarks demonstrate its effectiveness. In fairness-aware graph learning~\citep{MoE_4}, G-FAME and G-FAME++ plug MoE layers into GNNs and add diversity regularizers at node, layer, and expert levels to reduce redundancy and encourage distinct representations, yielding improved performance under fairness metrics. Ensemble learning has also been explored for malware detection. Reference~\citep{2_malware_stacking} proposes a recent stacking framework for Android malware that tackles class imbalance using a GAN together with Borderline-SMOTE, reduces redundant features through Information Gain followed by PCA, and employs an attention-based meta-learner to weight base classifiers. The resulting system improves generalization and stability on large-scale datasets.

As the need for interpretability has intensified, a growing literature has introduced explanation techniques tailored to GNNs. Although GNNs are effective at modeling structural patterns in program behavior, the layered message passing and nonlinear transformations make their decision processes opaque. To address this limitation, eXplainable Artificial Intelligence (XAI) methods such as GNNExplainer~\citep{GNNExplainer}, PGExplainer~\citep{PGExplainer}, SubgraphX~\citep{SubgraphX}, and CaptumExplainer~\citep{SAL_GBP} identify influential nodes, edges, or subgraphs that drive model outputs. In the context of CFGs, Herath et al.~\citep{CFGExplainer} proposed a model-agnostic pipeline that extracts influential subgraphs and ranks node importance using a surrogate model trained on node embeddings, thereby facilitating clear visualization of graph regions with the greatest impact on classification. Subsequent research~\citep{Consistency} presented a CFG–based malware detector that employs a hybrid node embedding combining rule-based encodings with autoencoder-derived features; after GNN-based classification, the framework applies multiple explainers, including GNNExplainer, PGExplainer, and Captum with several attribution strategies, and introduces RankFusion, an aggregation approach that fuses scores across explainers to yield more stable and informative attributions.

\section{Proposed Method} \label{sec:framework}

\begin{figure}
    \centering
    \includegraphics[width=\linewidth]{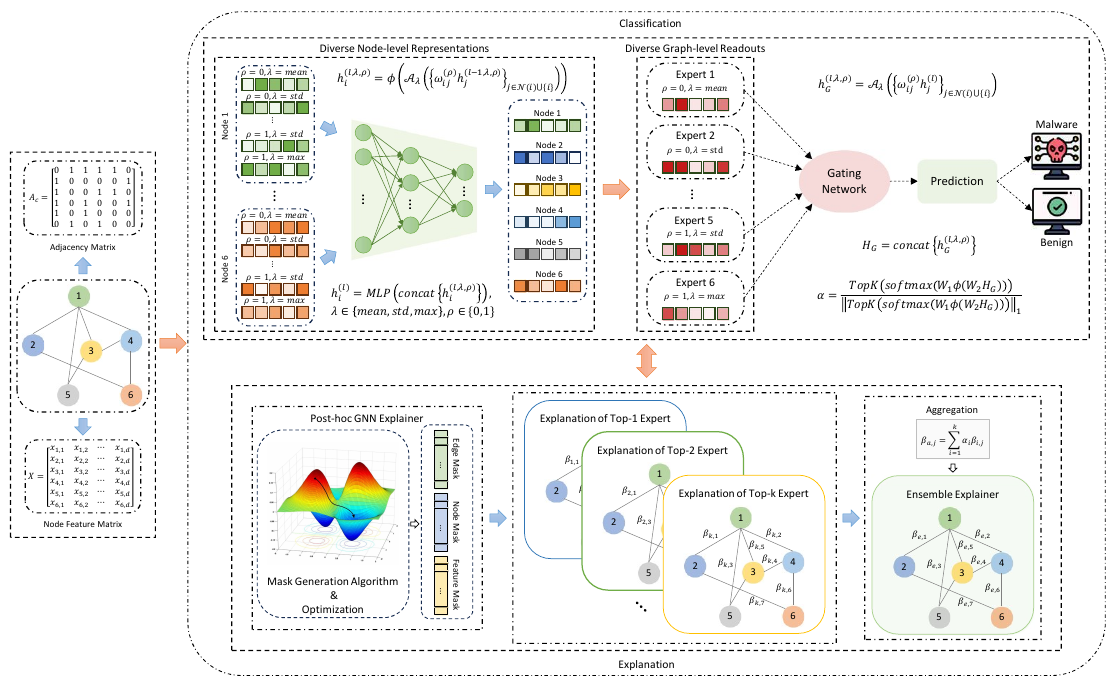}
    \caption{From CFG to prediction and explanation: diverse node representation, six experts, Top-k routing, and gate-weighted explanation.}
    \label{fig:framework}
\end{figure}

This section details our proposed framework and is organized as follows. We begin with CFG construction and feature representation, describing how CFGs are extracted from PE files and how basic-block feature vectors are built. We then introduce node-level multi-statistic encoding, a layerwise encoder that forms diverse node representations via degree reweighting and pooling choices. Next, graph-level expert readouts present six experts that aggregate node states into complementary graph embeddings. This is followed by top-k routing with LB, where a gating network selects and weights experts per input while encouraging balanced utilization. Finally, routing-aware post-hoc explanations outline how per-expert edge-level explanations are computed and combined with the learned gates to align the rationale with the model’s decisions. Figure \ref{fig:framework} provides an end-to-end overview of the proposed framework.

\subsection{CFG Construction and Feature Representation}

CFGs represent the execution semantics of a program, where each node corresponds to a basic block, which is a contiguous sequence of instructions with a single entry and a single exit point, and the directed edges capture the possible transitions between these blocks. While static CFGs derived from disassembly provide an abstract view of program structure, they often fail to capture certain execution paths introduced by obfuscation, indirect jumps, or dynamic code loading. In contrast, dynamic CFGs are constructed by monitoring a program’s runtime behavior, which exposes execution traces that static analysis may overlook. This dynamic perspective is particularly effective for analyzing sophisticated or evasive malware. In our framework, we employ dynamic symbolic execution for CFG extraction from PE files to obtain reliable and comprehensive graph representations for subsequent embedding and classification stages.

\begin{figure}
    \centering
    \includegraphics[width=\linewidth]{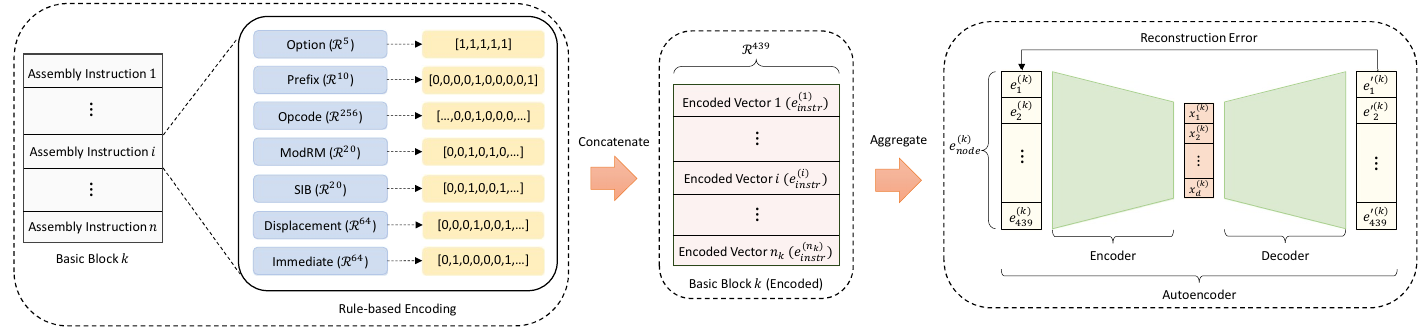}
    \caption{The two-step node feature embedding process, including rule-based instruction encoding and autoencoder-based dimensionality reduction.}
    \label{fig:embedding_pipeline}
\end{figure}

To prepare each node of the generated CFG for learning, we employ a two-stage node feature embedding pipeline, as illustrated in Figure~\ref{fig:embedding_pipeline}. The first stage encodes assembly instructions within each basic block using a rule-based scheme, and the second stage applies unsupervised dimensionality reduction through an autoencoder to generate compact and expressive node representations.

Each basic block consists of a sequence of x86-64 assembly instructions. Following a refined variant of the method proposed in~\citep{MalGNE}, each instruction is converted into a fixed-length feature vector using a comprehensive rule-based encoding process. This design ensures complete coverage of the x86-64 instruction set and mitigates the out-of-vocabulary issue that typically occurs in textual embeddings. Each instruction is decomposed into up to seven distinct components: prefix, opcode, ModRM, SIB, displacement, immediate, and an option flag. These components are encoded as follows:
\begin{enumerate}
    \item \textbf{Prefix:} This field includes four parts: extra segment (ES) register, operand-size override, address-size override, and lock prefix. The ES register can take seven possible values, while the remaining three are binary. Hence, the prefix is represented as a ten-dimensional one-hot vector.
    \item \textbf{Opcode:} The opcode defines the fundamental operation performed by the instruction and can assume up to 256 possible values. It is represented as a 256-dimensional one-hot vector.
    \item \textbf{ModRM:} The ModRM field is a single byte divided into mode (2 bits), register (3 bits), and memory address (3 bits), corresponding to 4, 8, and 8 possible values, respectively. This results in a 20-dimensional one-hot vector.
    \item \textbf{SIB:} The Scale-Index-Base (SIB) field is also a single byte divided into scale (2 bits), index (3 bits), and base (3 bits), which are encoded into a 20-dimensional one-hot vector.
    \item \textbf{Displacement:} The displacement represents an offset value used in memory addressing to calculate the effective address. It is encoded as a 64-dimensional binary vector.
    \item \textbf{Immediate:} The immediate value is a constant operand embedded directly within the instruction. It is encoded as a 64-dimensional binary vector.
    \item \textbf{Option:} Since the presence of prefix, ModRM, SIB, displacement, and immediate fields varies among instructions, a 5-dimensional binary vector is used to indicate their inclusion.
\end{enumerate}

Concatenating these components produces a 439-dimensional vector for each instruction, which preserves both operational semantics and operand structures.

Because a basic block may include several instructions, their encoded vectors are aggregated into a unified node representation:
\begin{equation}
{e}_{\text{node}}^{(k)} = \mathcal{A}\left( {e}_{\text{instr}}^{(1)}, {e}_{\text{instr}}^{(2)}, \ldots, {e}_{\text{instr}}^{(n_k)} \right)
\end{equation}
where \( n_k \) is the number of instructions in the \(k\)-th node (basic block), \( {e}_{\text{instr}}^{(i)} \in \mathbb{R}^{439} \) represents the encoded vector of the \(i\)-th instruction, and \( \mathcal{A}(\cdot) \) denotes an aggregation operator such as mean or max pooling. The resulting \( {e}_{\text{node}}^{(k)} \in \mathbb{R}^{439} \) forms the high-dimensional node embedding.

To further refine these representations, we train an autoencoder to compress the high-dimensional vectors while retaining their essential information. The autoencoder, composed of an encoder and decoder, is optimized using the mean squared error (MSE) loss:
\begin{equation}
L_{\text{MSE}} = \frac{1}{M} \sum_{i=1}^{M} \left\| {e}_{\text{instr}}^{(i)} - g_{\phi}(f_{\theta}({e}_{\text{instr}}^{(i)})) \right\|^2
\end{equation}
where \( f_{\theta} \) and \( g_{\phi} \) denote the encoder and decoder functions, respectively, and \( M \) is the number of instruction samples used during training.

After training, the encoder maps each node’s high-dimensional representation into a compact latent space:
\begin{equation}
{x}^{(k)} = f_{\theta}({e}_{\text{node}}^{(k)})
\end{equation}
where \( {x}^{(k)} \in \mathbb{R}^{d} \), with \( d \ll 439 \), represents the reduced-dimensional embedding of node \(k\).

This two-stage embedding process effectively balances expressiveness and compactness, which yields informative node features for downstream graph learning tasks, particularly in unsupervised or weakly supervised malware detection scenarios.

By collecting the embeddings from all nodes, we construct the node feature matrix \( {X} \in \mathbb{R}^{N \times d} \), where \( N \) is the total number of nodes in the CFG. Each row corresponds to one node embedding:
\begin{equation}
{X} = 
\begin{bmatrix}
{x}^{(1)} \\
{x}^{(2)} \\
\vdots \\
{x}^{(N)}
\end{bmatrix}
\in \mathbb{R}^{N \times d}
\end{equation}

\subsection{Node-Level Multi-Statistic Encoding}

GNNs are a category of neural architectures designed to process graph-structured data. In our context, the input is a CFG, where each node corresponds to a basic block represented by a feature vector \( {x}^{(i)} \in \mathbb{R}^d \), and edges encode control flow dependencies. GNNs operate through a mechanism known as message passing, wherein each node iteratively updates its representation by aggregating information from its neighboring nodes.

The general message passing framework for a GNN layer is expressed as:
\begin{equation}
{h}_i^{(l)} = \phi\left({h}_i^{(l-1)}, \mathcal{A}\left(\left\{ {h}_j^{(l-1)} \right\}_{j \in \mathcal{N}(i)}\right)\right)
\end{equation}
where \( {h}_i^{(l)} \) denotes the representation of node \( i \) at layer \( l \), \( \mathcal{N}(i) \) denotes the set of neighboring nodes of \( i \), and \(\phi(\cdot)\) is differentiable nonlinear function (e.g., ReLU). At the initial layer, \( {h}_i^{(0)} = {x}^{(i)} \), the input node feature vector.

We design a node representation that explicitly promotes diversity at the node level by running multiple, complementary neighborhood statistics in parallel and then fusing them. Let the input be a CFG with node features \({x}^{(i)}\in\mathbb{R}^d\). At layer \(l\), each node \(i\) aggregates messages from its closed neighborhood \(\mathcal{N}(i)\cup\{i\}\) using degree-aware weights and one of several pooling operators. The per-layer update is:
\begin{equation}    
{h}_i^{(l,\lambda,\rho)}
=\phi\left(
\mathcal{A}_{\lambda}\left(
\left\{ \omega_{ij}^{(\rho)}\,{h}_j^{(l-1,\lambda,\rho)}\right\}_{j\in \mathcal{N}(i)\cup\{i\}}
\right)\right)
\end{equation}
where \(\lambda\in\{\text{mean},\text{std},\text{max}\}\) selects the pooling statistic, and \(\rho\in\{0,1\}\) controls whether aggregation is uniform or degree-reweighted.

Let \(d_j\) denote the degree of node \(j\) in the current graph, and let
\begin{equation}
\tilde{w}_{ij}^{(\rho)}
=\begin{cases}
1, & \rho=0\\[2pt]
d_j, & \rho=1
\end{cases}
\qquad\text{and}\qquad
\omega_{ij}^{(\rho)}
=\dfrac{\tilde{w}_{ij}^{(\rho)}}{\sum\limits_{k\in \mathcal{N}(i)\cup\{i\}} \tilde{w}_{ik}^{(\rho)}}\,
\end{equation}

When \(\rho=0\), all neighbors (and the self term) contribute equally, which yields a uniform aggregation that does not depend on node connectivity. When \(\rho=1\), messages from higher-degree nodes receive larger weight, biasing the aggregation toward structurally central neighbors while still normalizing so \(\sum_j \omega_{ij}^{(\rho)}=1\).

Given the weighted messages \({m}_{ij}=\omega_{ij}^{(\rho)}{h}_j^{(l-1,\lambda,\rho)}\), we define mean, max, and standard deviation aggregation (std) as:
\begin{equation}
\mathcal{A}_{\text{mean}}\left(\{{m}_{ij}\}\right) =\sum\nolimits_{j}{m}_{ij}
\label{eq:mean}
\end{equation}

\begin{equation}
    \mathcal{A}_{\text{max}}\left(\{{m}_{ij}\}\right) =\max\nolimits_{j}\,{m}_{ij}\;\,
    \label{eq:max}
\end{equation}

\begin{equation}
\mathcal{A}_{\text{std}}\!\left(\{{m}_{ij}\}\right) =\sqrt{\Big(\sum\nolimits_{j}{m}_{ij}\odot {m}_{ij}\Big) - {\mu}_i\odot {\mu}_i}\,,\quad
{\mu}_i=\sum\nolimits_{j}{m}_{ij}
\label{eq:std}
\end{equation}
A large standard deviation response indicates high dispersion among neighbors along that feature dimension, while a small response indicates a homogeneous neighborhood.

We build a diverse node representation by concatenating the six channels produced by \(\rho\in\{0,1\}\) and \(\lambda\in\{\text{mean},\text{std},\text{max}\}\), then applying a lightweight MLP:
\begin{equation}
{h}_i^{(l)}
=\mathrm{MLP}\left(\operatorname{concat}\{{h}_i^{(l,\lambda,\rho)}\;:\;\lambda\in\{\text{mean},\text{std},\text{max}\},\;\rho\in\{0,1\}\}\right)
\end{equation}
This construction yields complementary views of each node’s neighborhood that are later exploited by expert-specific readouts and the router.

\subsection{Graph-Level Expert Readouts}

In an MoE, each expert is a simple predictor trained to specialize on a particular view of the input. Rather than relying on one universal readout, the model exposes multiple, complementary readouts and later lets a router weight them per graph. This promotes specialization and reduces interference across heterogeneous program behaviors.

We instantiate six graph experts, each tied to a specific combination of a degree weighting choice \(\rho\in\{0,1\}\) and a pooling statistic \(\lambda\in\{\text{mean},\text{std},\text{max}\}\). The degree weighting \(\rho=0\) corresponds to a uniform prior where all nodes contribute equally, while \(\rho=1\) corresponds to a degree-biased prior where higher-degree nodes receive proportionally larger weight. The pooling statistic \(\lambda\) selects the aggregation function defined in \eqref{eq:mean}–\eqref{eq:std}: mean captures central tendency, standard deviation captures dispersion, and max highlights extreme responses.

Formally, the \((\rho,\lambda)\) expert forms a graph-level representation by aggregating the final-layer node representations \(\{{h}_j^{(l)}\}\) with degree-aware weights \(\omega_{ij}^{(\rho)}\):
\begin{equation}
\label{eq:expert-readout}
{h}_{G}^{(l,\lambda,\rho)}
=
\mathcal{A}_{\lambda}\!\left(
\left\{
\,\omega_{ij}^{(\rho)}\,{h}_{j}^{(l)}
\right\}_{j\in \mathcal{N}(i)\cup\{i\}}
\right)
\end{equation}
where \(\mathcal{A}_{\lambda}\) is one of the three statistics in \eqref{eq:mean}–\eqref{eq:std}, and \(\omega_{ij}^{(\rho)}\) encodes the uniform or degree-biased weighting induced by \(\rho\) (normalized across the aggregation set).

Crossing the two structural priors with the three pooling statistics yields six experts, each producing class logits from its designated \((\rho,\lambda)\) readout. These experts provide complementary graph-level views that capture distinct structural and statistical aspects of the CFG. In the next stage, a gating network combines the six expert logits into a single decision by assigning input-dependent weights to the experts.

\subsection{Top-k Routing with Load Balancing}

The gating network receives a single graph descriptor obtained by concatenating the six expert
readouts:
\begin{equation}
\label{eq:Hg_concat}
{H}_G \;=\; \operatorname{concat}\!\left\{\,{h}_G^{(\lambda,\rho)}\,\right\}_{\rho\in\{0,1\},\,\lambda\in\{\mathrm{mean},\mathrm{std},\mathrm{max}\}}
\;\in\mathbb{R}^{6H}.
\end{equation}
A two-layer MLP maps ${H}_G$ to six routing logits; these are converted to probabilities and then sparsified by a Top-$k$ operation, after which the surviving entries are renormalized to sum to one:
\begin{equation}
\label{eq:alpha_gate}
\boldsymbol{\alpha}
\;=\;
\frac{\operatorname{TopK}\!\big(\operatorname{softmax}\!\big({W}_1\,\phi({W}_2\,{H}_G)\big)\big)}
{\big\lVert \operatorname{TopK}\!\big(\operatorname{softmax}\!\big({W}_1\,\phi({W}_2\,{H}_G)\big)\big)\big\rVert_1}
\;\in\mathbb{R}^{6},
\end{equation}
where ${W}_2:\mathbb{R}^{6H}\!\to\!\mathbb{R}^{H}$, ${W}_1:\mathbb{R}^{H}\!\to\!{R}^{6}$, and $\phi(\cdot)$ is a pointwise nonlinearity.
Concretely, $\operatorname{TopK}(\cdot)$ keeps only the $k$ largest probabilities (sets the other $6-k$ to zero).
Dividing by the $\ell_1$ norm of that sparse vector renormalizes the kept entries so that
$\sum_{e=1}^{6}\alpha_e=1$ and exactly $k$ of them are nonzero. Thus, the gate ${\alpha}$ is a sparse,
input-dependent mixture over experts.

Let ${o}_e\in\mathbb{R}^{2}$ be the class-logit vector produced by expert $e$. The final logits are
\begin{equation}
\label{eq:final_logits}
\hat{{y}} \;=\; \sum_{e=1}^{6} \alpha_e\,{o}_e,
\end{equation}
followed by a softmax for prediction at inference and cross-entropy during training.

To discourage expert collapse, we include a load balancing auxiliary loss computed from the
batch-average gate $q_e=\frac{1}{B}\sum_{b=1}^{B}\alpha_{e}^{(b)}$:
\begin{equation}
\label{eq:lb_loss}
\mathcal{L}_{\mathrm{LB}} \;=\; \sum_{e=1}^{6} q_e \,\log(6\,q_e),
\end{equation}
which encourages more uniform utilization across experts. The total objective is
$\mathcal{L}=\mathcal{L}_{\mathrm{CE}}+\lambda_{\mathrm{LB}}\,\mathcal{L}_{\mathrm{LB}}$ where \(\lambda_{\mathrm{LB}}\) is a tuning coefficient, and \(\mathcal{L}_{\mathrm{CE}}\) denotes the binary cross-entropy classification loss.

\subsection{Routing-Aware Post-Hoc Explanation}

We seek explanations that reflect both what each expert highlights and how strongly the router relies on that expert for a given graph. Let the gating network produce nonnegative, normalized weights \(\{\alpha_i\}_{i=1}^{k}\) over the Top-\(k\) selected experts for an input graph. For each selected expert \(i\), we run a post-hoc graph explainer (e.g., gradient-based attribution, perturbation or masking, or a local surrogate) on the expert’s prediction head to obtain per-edge importance scores \(\beta_{(i,j)} \) for edge \(j\) in the current graph. We optionally normalize each \(\beta_{(i,\cdot)}\) across edges to a common scale to improve comparability across experts.

We then form a routing-aware aggregated edge score by weighting the per-expert scores with the router gates:
\begin{equation}
\label{eq:router_aware_agg}
\beta_{(a,j)} \;=\; \sum_{i=1}^{k} \alpha_i \, \beta_{(i,j)} \,,
\end{equation}
where \(\beta_{(a,j)}\) is the final explanation score for edge \(j\). This aggregation aligns the explanation with the model’s actual combination of experts for the specific input. Crucially, because routing selects only \(k\) experts, we compute explanations for those \(k\) experts rather than for all six; this reduces cost and keeps the explanation faithful to the model’s conditional computation.

\subsection{Evaluation Metrics}
To comprehensively assess both the classification performance and interpretability of the proposed framework, we employ the following evaluation metrics:

\paragraph{Accuracy}
Accuracy measures the proportion of correctly classified samples over the total number of samples:
\begin{equation}
\text{Accuracy:} = \frac{TP + TN}{TP + TN + FP + FN}
\end{equation}
where \( TP \), \( TN \), \( FP \), and \( FN \) represent the number of true positives, true negatives, false positives, and false negatives, respectively.

\paragraph{Precision:}
Precision evaluates the correctness of positive predictions, defined as:
\begin{equation}
\text{Precision} = \frac{TP}{TP + FP}
\end{equation}
A high precision indicates a low false positive rate.

\paragraph{Recall:}
Recall (or sensitivity) assesses the model's ability to identify all relevant instances:
\begin{equation}
\text{Recall} = \frac{TP}{TP + FN}
\end{equation}
A high recall reflects a low false negative rate, which is particularly important in malware detection.

\paragraph{F1 score:}
The F1 score is the harmonic mean of precision and recall, balancing the trade-off between the two:
\begin{equation}
\text{F1 score} = 2 \cdot \frac{\text{Precision} \cdot \text{Recall}}{\text{Precision} + \text{Recall}}
\end{equation}

\paragraph{Fidelity:}
Fidelity quantifies the explanatory power of a subgraph by assessing its influence on the model's prediction. It measures how the prediction outcome changes when either the important or unimportant portions of the graph are removed. Fidelity is evaluated in two complementary forms:
\begin{equation}
    Fidelity{+} = 1 - \frac{1}{N} \sum_{i=1}^{N} \mathds{1}\left( \hat{y}_i^{G_{C \setminus S}} = \hat{y}_i \right),
\end{equation}

\begin{equation}
    Fidelity{-} = 1 - \frac{1}{N} \sum_{i=1}^{N} \mathds{1}\left( \hat{y}_i^{G_{S}} = \hat{y}_i \right),
\end{equation}
where \( G_S \) denotes the important subgraph, \( G_C \) is the complete original graph, \( \hat{y}_i \) is the model's predicted label for the \( i \)-th sample, and \( y_i \) is the ground truth label.

$Fidelity{+}$ captures the impact of the removed important subgraph by comparing the model’s predictions on the original graph and on the graph with the important part removed (\( G_{C \setminus S} \)). A higher $Fidelity{+}$ indicates that the removed subgraph had a substantial influence on the model’s decision, thereby validating its importance in the prediction process.
Conversely, $Fidelity{-}$  assesses the predictive sufficiency of the important subgraph alone by comparing the model’s prediction on the full graph and on \( G_S \). A lower $Fidelity{-}$ implies that the identified subgraph retains most of the critical information needed for accurate prediction, thus indicating a more faithful and self-contained explanation.
Together, these two metrics provide a comprehensive evaluation of explanation quality by quantifying both the necessity and sufficiency of the identified subgraph in relation to the model’s output.

\paragraph{Characterization:}
To provide a unified assessment of explanation quality, we adopt the Characterization score (Char) proposed in~\citep{char}. 
This metric integrates the two complementary fidelity measures, $Fidelity{+}$ and $Fidelity{-}$, to evaluate both the \textit{necessity} and \textit{sufficiency} of an explanation. 
While $Fidelity{+}$ quantifies how critical the identified subgraph is to the model’s prediction (necessity), $(1 - Fidelity{-})$ measures how well the subgraph alone can reproduce the model’s prediction (sufficiency). 
The Characterization score combines these two aspects using a weighted harmonic mean:
\begin{equation}
\text{Char} = 
\frac{(w_{+} + w_{-}) \times Fidelity{+} \times (1 - Fidelity{-})}
{w_{+} \times (1 - Fidelity{-}) + w_{-} \times Fidelity{+}}
\end{equation}
where $w_{+}$ and $w_{-}$ are the respective weights assigned to $Fidelity{+}$ and $(1 - Fidelity{-})$, satisfying $w_{+} + w_{-} = 1$. 
Setting equal weights ($w_{+} = w_{-} = 0.5$) yields a balanced evaluation of both necessity and sufficiency. 
A high Characterization score indicates that the explanation subgraph is simultaneously indispensable and self-sufficient in determining the model’s prediction, thus providing a more comprehensive measure of explainability quality.

\section{Results and Analysis}
\label{sec:result}

In our experiments, malicious samples were drawn from the BODMAS~\citep{yang2021bodmas} and PMML~\citep{practicalsecurity2024pe} datasets, while benign samples were sourced from DikeDataset~\citep{dikedataset}. An overview of all datasets used appears in Table~\ref{tab:dataset_stats}. For CFG extraction, we employed the angr framework~\citep{shoshitaishvili2016state}, a Python-based binary analysis toolkit that constructs CFGs by coupling symbolic execution with constraint solving, enabling precise and granular graph recovery. All experiments were executed on a workstation with an Intel Xeon Platinum 8253 CPU (32 cores at 3.0\,GHz) and 128\,GB RAM. The framework was implemented in Python, leveraging PyTorch Geometric for model training and NetworkX~v2.8.8 for graph processing and manipulation.

\begin{table}
\centering
\caption{Characteristics of the evaluated datasets.}
\label{tab:dataset_stats}
\begin{tabular}{l*{4}{c}}
\toprule
\textbf{Dataset} & \textbf{\#Samples} & \textbf{Avg. Nodes} & \textbf{Avg. Edges} & \textbf{Label} \\
\midrule
BODMAS      & 122 & 63{,}226.44 & 66{,}033.72 & Malware \\
DikeDataset & 319 & 9{,}059.07  & 15{,}171.37 & Benign  \\
PMML        & 390 & 14{,}246.54 & 23{,}977.81 & Malware \\
\bottomrule
\end{tabular}%
\end{table}

To compress the original 439\,-dimensional node features, we trained a symmetric autoencoder that maps inputs to a 64\,-dimensional latent space. The encoder comprises three fully connected layers (439$\!\rightarrow$256$\!\rightarrow$128$\!\rightarrow$64) with ReLU activations; the decoder mirrors this topology in reverse (64$\!\rightarrow$128$\!\rightarrow$256$\!\rightarrow$439), also with ReLU. We optimized mean squared error (MSE) reconstruction loss using Adam (learning rate $1\times10^{-4}$) for 5{,}000 epochs. Training was stopped at 5{,}000 epochs as the validation MSE had stabilized below $1\times10^{-4}$ for the final 1{,}000 epochs, indicating convergence. The 64\,-dimensional encoder outputs were subsequently used as node feature embeddings for downstream graph learning.

To evaluate the effectiveness of the proposed mixture of experts framework, we compare against two recent state-of-the-art graph based malware detection frameworks that have been studied on the same dataset explained in Table~\ref{tab:dataset_stats}.
First, we include MalGNE~\citep{MalGNE}, which employs a dedicated graph embedding strategy combined with standard message passing backbones. Following the original design, we evaluate MalGNE with both Graph Convolutional Network (GCN)~\citep{GCN} and Graph Attention Network (GAT)~\citep{GAT} variants.
Second, we include the consistency driven graph learning framework~\citep{Consistency}. In alignment with that work, we evaluate four backbone instantiations: GCN, GAT, GIN, and GraphSAGE.

To explore expert aggregation strategies, five MoE variants were implemented: (i) MoE with uniform aggregation, where all experts contribute equally; (ii) MoE with temperature softmax, which assigns adaptive expert weights through a temperature-scaled softmax function; (iii) MoE with Top-1 expert selection, activating only the most relevant expert for each input; (iv) MoE with Top-2 expert selection without LB, which omits the regularization that enforces balanced expert utilization, and (v) MoE with Top-2 expert selection with LB.

In MoE with temperature softmax variant, the router assigns a dense weight to every expert rather than selecting a Top-$k$ subset. Let ${s}\in\mathbb{R}^6$ be the router logits computed from the graph representation. The gate for expert $e$ is
\begin{equation}
p_e \;=\; \frac{\exp\!\big(s_e / T\big)}{\sum_{j=1}^{6} \exp\!\big(s_j / T\big)}\,,
\end{equation}
with a fixed temperature $T=0.5$ in our implementation. Lower $T$ sharpens the distribution, higher $T$ smooths it. This scenario keeps all experts active on every input while adjusting their contributions according to the router’s confidence.

All scenarios were trained under a common configuration to ensure fair comparison. The dataset was stratified and split into 80\% for training and 20\% for testing, preserving equal class proportions for benign and malicious samples in both subsets. We used a mini-batch size of $8$, which balances GPU memory constraints with gradient stability on graph-structured inputs. Each model was optimized for $100$ epochs. The node encoder employed a hidden width of $64$ channels and a depth of $3$ layers, providing sufficient capacity to capture multi-hop program structure. We applied dropout at rate $0.2$ after intermediate transformations to regularize representations. Optimization used Adam with a learning rate of $3\times10^{-4}$, which we found to yield stable convergence across GNN backbones and MoE variants. For MoE scenarios, a LB auxiliary term was included with coefficient $\lambda_{LB}=0.01$ to encourage even expert utilization; this term was disabled in the ``without LB'' ablation. All remaining settings were held fixed across scenarios to isolate the impact of routing and expert configuration.

Table~\ref{tab:classwise_results} reports classwise precision, recall, F1, and overall accuracy for all models. In both comparison frameworks, single-backbone models achieve accuracies between 85.90\% and 89.10\%, with the best baseline result obtained by the GAT backbone under the MalGNE framework and GraphSAGE backbone under the Consistensy framework at 89.10
All MoE variants surpass these baselines. Even without routing, uniform aggregation improves accuracy to 92.31\%, highlighting the benefit of combining diverse graph-level representations. Temperature-based softmax achieves 89.10\%, while Top-1 routing increases performance to 91.03\%, providing balanced precision and recall across both classes.
Top-2 routing further strengthens performance. Without load balancing, accuracy reaches 91.67\%, whereas incorporating the load-balancing term yields the best overall result at 93.59\%. This configuration achieves the highest benign precision at 96.23\% and the highest malicious recall at 97.94\%, resulting in the strongest F1 scores for both benign at 91.07\% and malicious at 95\%.
The comparison between Top-2 with and without load balancing confirms that regularized expert utilization improves stability and classwise balance. Overall, conditional multi-expert aggregation with controlled routing provides more reliable detection and better per-class trade-offs than single-backbone configurations.

\begin{table}
\centering
\caption{Classwise performance of different models.}
\label{tab:classwise_results}
\begin{tabular}{lccc ccc c}
\toprule
\multirow{2}{*}{Model} & \multicolumn{3}{c}{Benign} & \multicolumn{3}{c}{Malicious} & \multirow{2}{*}{Accuracy} \\
\cmidrule(lr){2-4}\cmidrule(lr){5-7}
 & Precision & Recall & F1 & Precision & Recall & F1 & \\
\midrule

\multicolumn{8}{l}{\textit{MalGNE framework~\citep{MalGNE}}} \\
GCN backbone & 81.97 & 84.75 & 83.33 & 90.53 & 88.66 & 89.58 & 87.18 \\
GAT backbone & 86.21 & 84.75 & 85.47 & 90.82 & 91.75 & 91.28 & 89.10 \\
\midrule

\multicolumn{8}{l}{\textit{Consistency framework~\citep{Consistency}}} \\
GCN backbone & 81.67 & 83.05 & 82.35 & 89.58 & 88.66 & 89.12 & 86.54 \\
GAT backbone & 81.97 & 84.75 & 83.33 & 90.53 & 88.66 & 89.58 & 87.18 \\
GIN backbone & 86.64 & 77.97 & 80.70 & 87.13 & 90.72 & 88.89 & 85.90 \\
GraphSAGE backbone & 85.00 & 86.44 & 85.71 & 91.67 & 90.72 & 91.19 & 89.10 \\
\midrule

\multicolumn{8}{l}{\textit{Proposed MoE variants}} \\
MoE-Uniform Aggregation  & 92.73 & 86.44 & 89.47 & 92.08 & 95.88 & 93.94 & 92.31 \\
MoE-Temperature Softmax & 83.87 & 88.14 & 85.95 & 92.55 & 89.69 & 91.10 & 89.10 \\
MoE-Top-1 Expert  & 86.89 & \textbf{89.83} & 88.33 & \textbf{93.68} & 91.75 & 92.71 & 91.03 \\
MoE-Top-2 Experts w/o LB & 92.59 & 84.75 & 88.50 & 91.18 & 95.88 & 93.47 & 91.67 \\
MoE-Top-2 Experts & \textbf{96.23} & 86.44 & \textbf{91.07} & 92.23 & \textbf{97.94} & \textbf{95.00} & \textbf{93.59} \\
\bottomrule
\end{tabular}
\end{table}

\begin{figure}
    \centering
    \begin{subfigure}{0.48\linewidth}
        \centering
        \includegraphics[width=\linewidth]{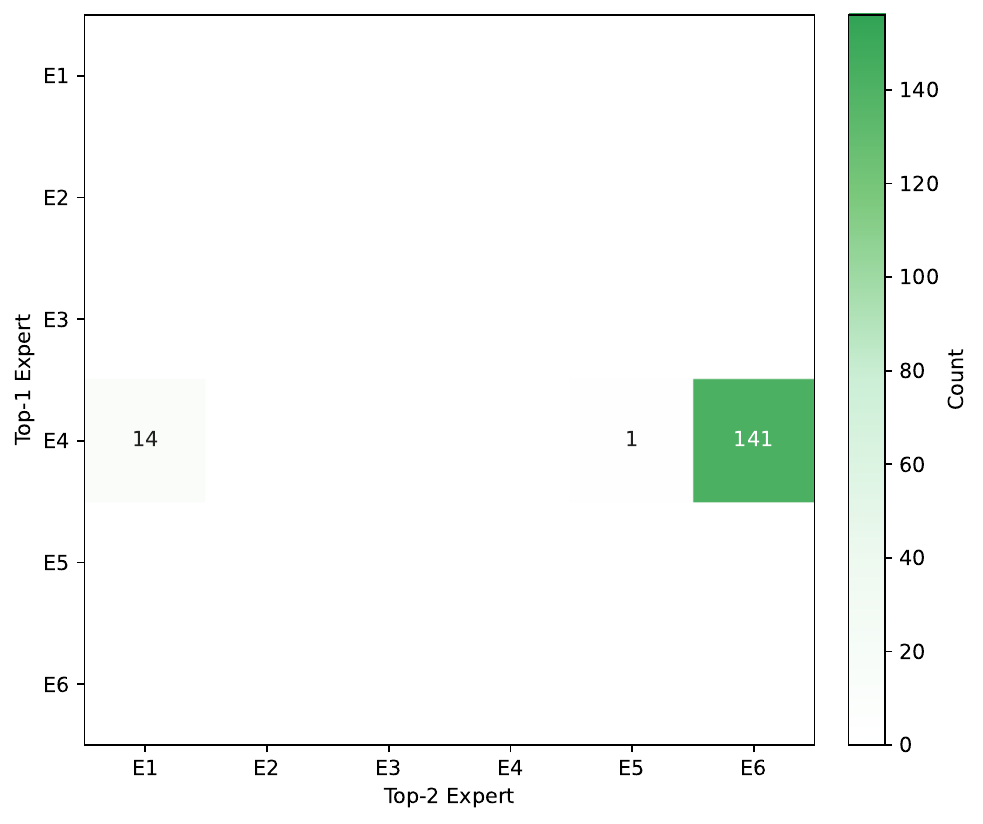}
        \caption{Without Load Balancing}
        \label{fig:coselection-nolb}
    \end{subfigure}\hfill
    \begin{subfigure}{0.48\linewidth}
        \centering
        \includegraphics[width=\linewidth]{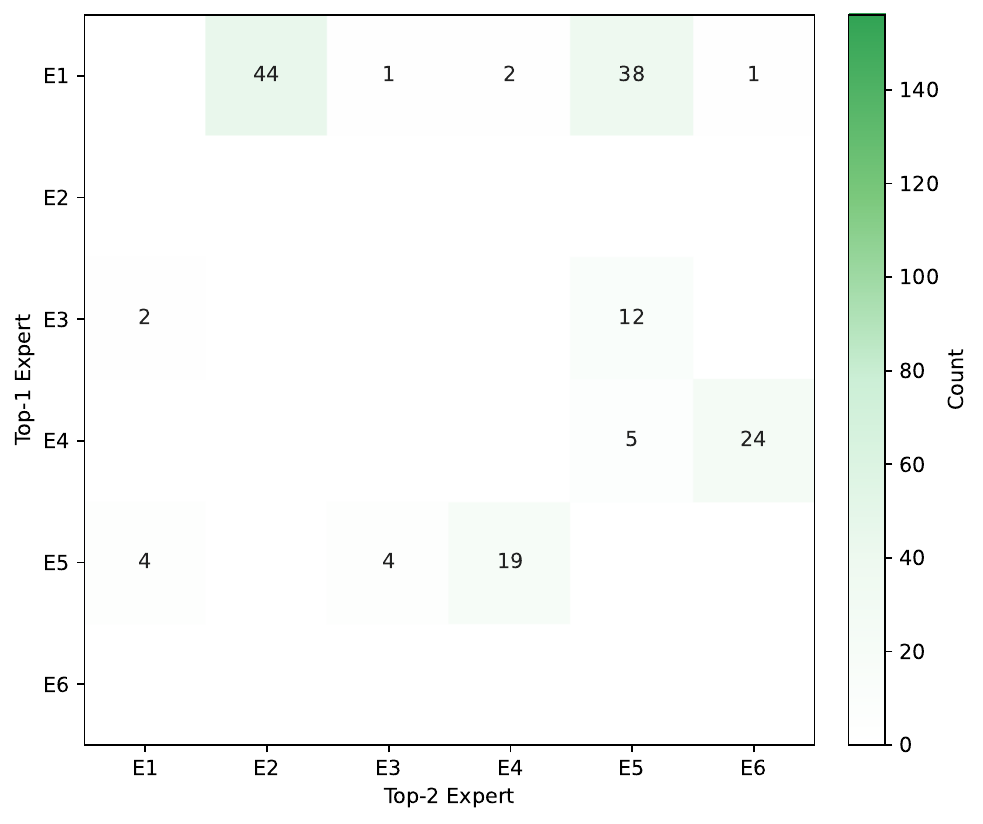}
        \caption{With Load Balancing}
        \label{fig:coselection-lb}
    \end{subfigure}
    \caption{Expert co-selection analysis in the Top-2 MoE configuration, comparing the diversity of expert activation {with} and {without} the load balancing term.}
    \label{fig:coselection-combined}
\end{figure}

\begin{figure}
    \centering
    \includegraphics[width=0.75\linewidth]{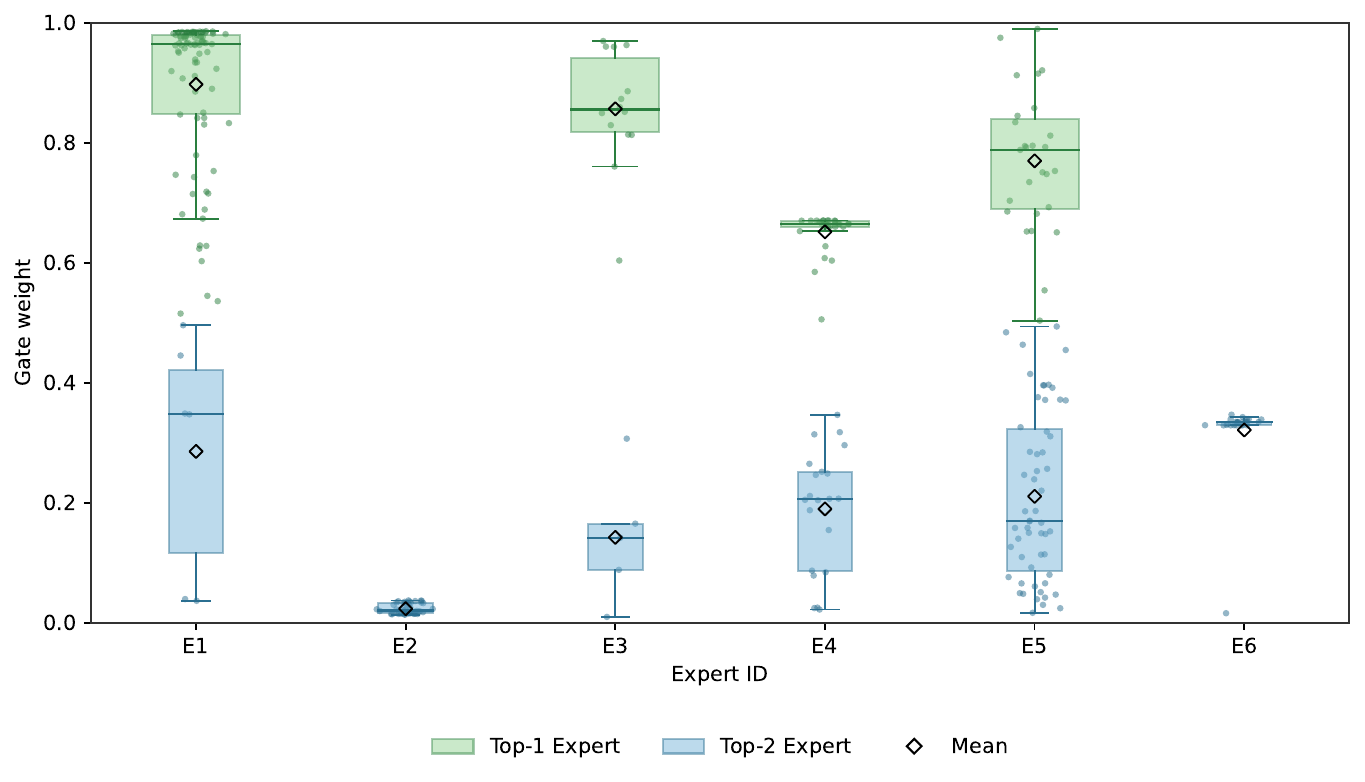}
    \caption{Distribution of gating weights across experts in the Top-2 MoE configuration with load balancing.}
    \label{fig:expert-weights}
\end{figure}

Figure~\ref{fig:coselection-combined} illustrates the expert co-selection patterns for the Top-2 MoE model, comparing runs conducted with and without the LB term in the objective function. Each heatmap shows the frequency with which a pair of experts was jointly activated during the test phase, where the $y$-axis denotes the Top-1 expert and the $x$-axis the Top-2 expert. As observed in Figure~\ref{fig:coselection-nolb}, removing the LB constraint leads to a strong selection bias toward a limited subset of expert pairs, with most test samples dominated by the combination of $(E4, E6)$. In contrast, the inclusion of the LB term (Figure~\ref{fig:coselection-lb}) significantly increases the diversity of expert utilization, promoting more balanced participation across all six experts. This diversity indicates that the LB regularization effectively mitigates expert over-specialization and encourages cooperative contribution among the expert modules. The experts are defined as follows: E1 ($\rho=0,\lambda=\text{mean}$), E2 ($\rho=0,\lambda=\text{std}$), E3 ($\rho=0,\lambda=\text{max}$), E4 ($\rho=1,\lambda=\text{mean}$), E5 ($\rho=1,\lambda=\text{std}$), and E6 ($\rho=1,\lambda=\text{max}$).
Moreover, Figure~\ref{fig:expert-weights} presents the distribution of gating weights assigned to each expert during the test phase in the Top-2 MoE configuration with LB. The green boxplots correspond to the Top-1 expert weights, while the blue boxplots represent the Top-2 expert weights across all six experts. The results reveal that the gating network dynamically allocates varying degrees of importance to different experts, indicating adaptive specialization rather than uniform weighting. In particular, E1 associated with distinct parameter settings, $(\rho=0,\lambda=\text{mean})$, tends to receive higher average weights, reflecting its stronger contribution to the final decision process. Overall, the distribution confirms that the LB term encourages equitable yet selective expert utilization, preventing dominance by a single expert.

\begin{figure}
    \centering
    \includegraphics[width=0.5\linewidth]{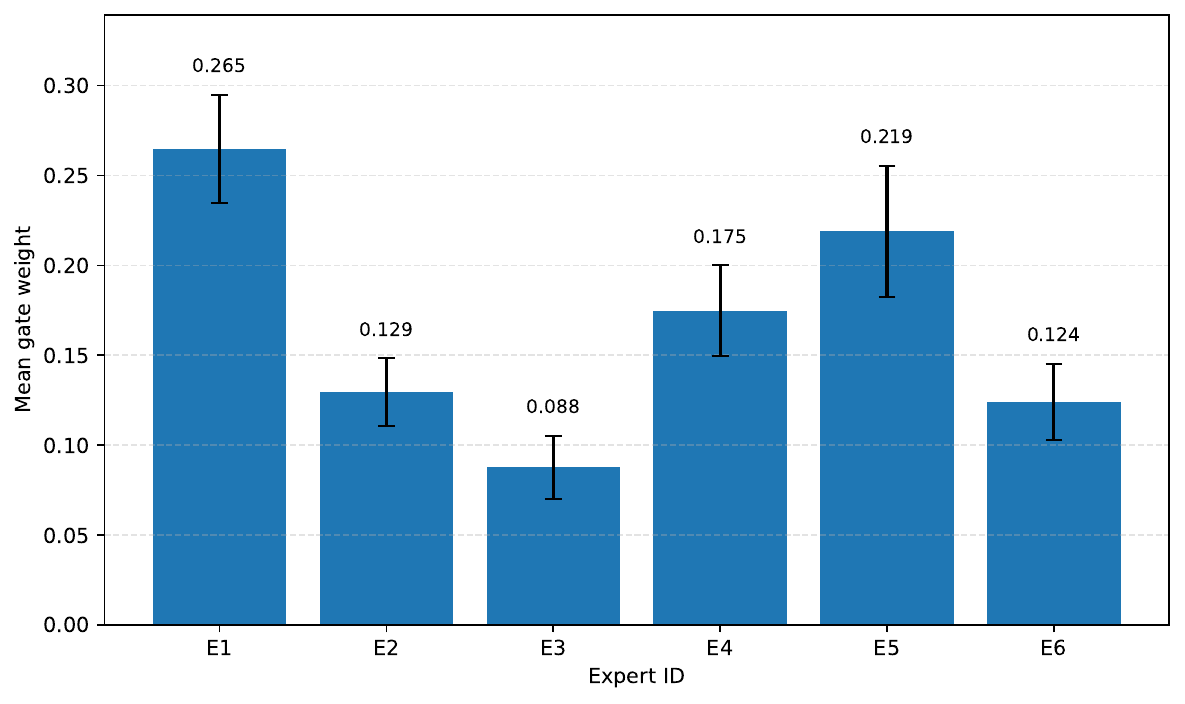}
    \caption{Average gate weights assigned to each expert in the temperature softmax configuration with 95\% confidence intervals.}
    \label{fig:tempsoftmax-mean-ci}
\end{figure}

\begin{figure}
    \centering
    \includegraphics[width=0.5\linewidth]{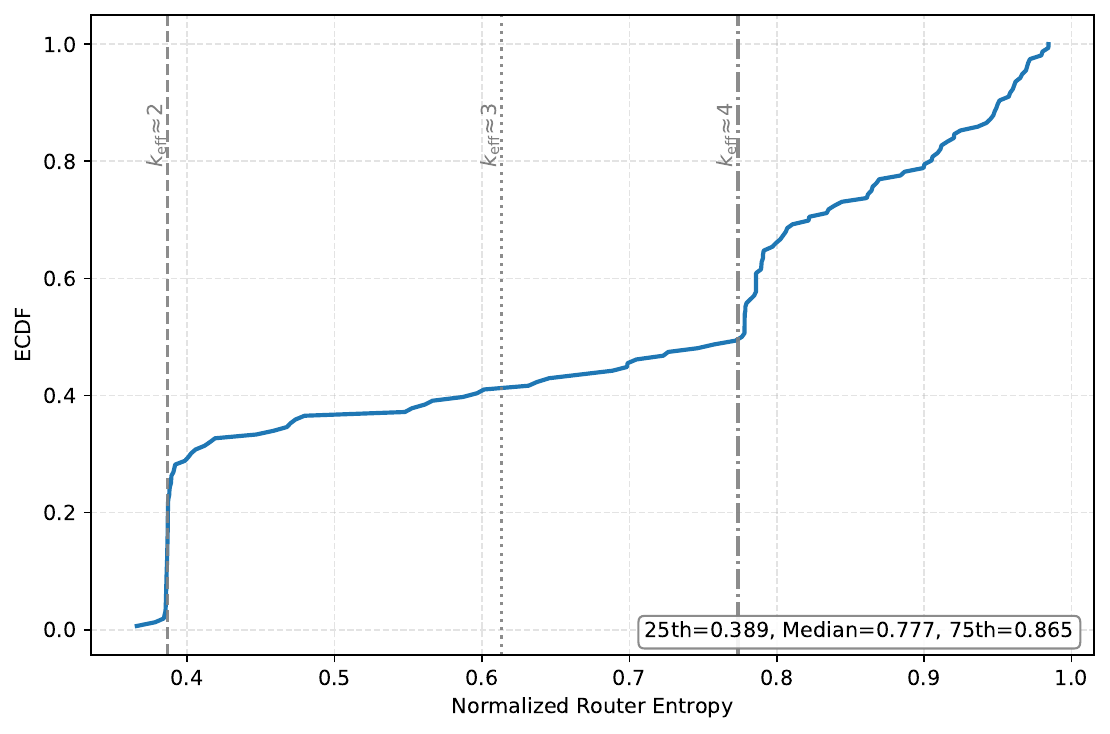}
    \caption{ECDF of normalized router entropy \(\tilde{H}(\boldsymbol{\alpha})\) in the temperature softmax configuration.}
    \label{fig:tempsoftmax-entropy-ecdf}
\end{figure}

Figure~\ref{fig:tempsoftmax-mean-ci} reports per-expert routing statistics for the MoE temperature softmax configuration on the test set. Bars indicate the mean gate weight assigned to each expert and the whiskers show 95\% confidence intervals. The router assigns larger average mass to certain experts (notably E1 and E5), while other experts receive smaller shares, indicating partial specialization under temperature-controlled soft weighting.
Moreover, Figure~\ref{fig:tempsoftmax-entropy-ecdf} characterizes the distribution of router uncertainty via the empirical cumulative distribution function (ECDF) of the normalized entropy of the gate vectors. Let the router output a gate (routing) distribution \({\alpha} = (\alpha_1,\ldots,\alpha_6)\), where \(\alpha_i \ge 0\) and \(\sum_{i=1}^{6}\alpha_i = 1\). The normalized router entropy is
\begin{equation}
\tilde{H}(\boldsymbol{\alpha}) \;=\; \frac{-\sum_{i=1}^{6} \alpha_i \,\log \alpha_i}{\log 6} \;\in\; [0,1],
\end{equation}
with values near \(0\) indicating peaked routing (one expert dominates) and values near \(1\) indicating diffuse routing (weights spread across many experts). Over the test set of size \(N\), the ECDF of \(\tilde{H}\) is
\begin{equation}
\mathrm{ECDF}(t) \;=\; \frac{1}{N}\,\bigl|\{\, s \in \{1,\ldots,N\} : \tilde{H}_s \le t \,\}\bigr|, \qquad t \in [0,1],
\end{equation}
where \(t\) is the entropy threshold on the horizontal axis and \(N\) is the number of test samples. The vertical reference lines in Figure~\ref{fig:tempsoftmax-entropy-ecdf} mark the normalized entropies associated with approximately \(k\) equally weighted experts,
\begin{equation}
\tilde{H}_k \;=\; \frac{\log k}{\log 6} \quad \text{for} \quad k \in \{2,3,4\},
\end{equation}
which provide visual anchors for interpreting how concentrated the routing is. For example, samples to the left of \(\tilde{H}_2\) are more concentrated than a two-expert split, while samples to the right of \(\tilde{H}_4\) are more diffuse than a four-expert split. The numbers reported in the bottom-right summary box (25th, Median, 75th) are the first quartile, median, and third quartile of the \(\tilde{H}\) distribution across test samples, respectively. They indicate that 25\% of samples have \(\tilde{H}\) at or below the first quartile, 50\% at or below the median, and 75\% at or below the third quartile. Compared with the Top-2 Expert MoE with LB, the temperature softmax variant typically yields higher \(\tilde{H}\) values, reflecting more diffuse allocation of mass across experts without enforcing pairwise selection.

\begin{figure}
    \centering
    \includegraphics[width=0.5\linewidth]{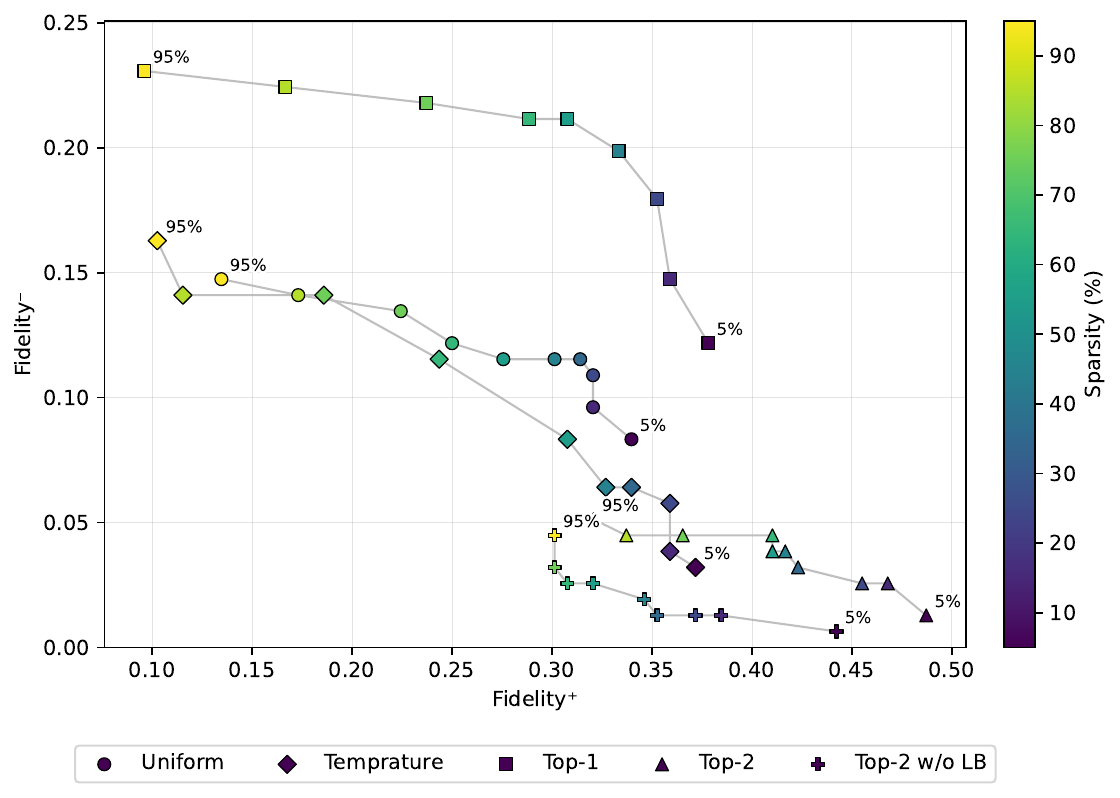}
    \caption{Comparison of $Fidelity^{+}$ and $Fidelity^{-}$ across graph sparsity levels for five MoE configurations.}
    \label{fig:pareto_fidelity}
\end{figure}

\begin{figure}
    \centering
    \includegraphics[width=0.5\linewidth]{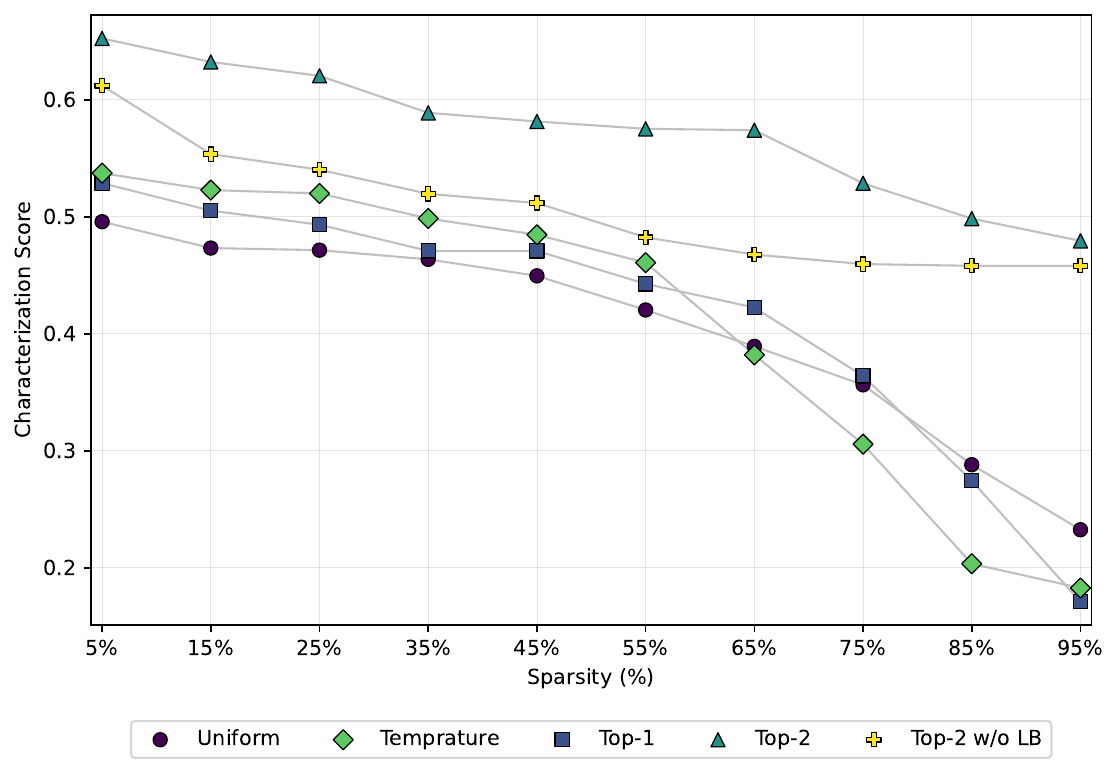}
    \caption{Characterization Score across sparsity levels for the evaluated MoE configurations.}
    \label{fig:characterization_sparsity}
\end{figure}

The final set of experiments evaluates the explainability performance of the proposed framework across five MoE configurations: uniform aggregation, temperature softmax, Top-1 Expert, Top-2 Expert, and Top-2 Expert without LB. Integrated Gradients from the Captum library~\citep{Captum} was employed to assign edge-level importance scores and quantify the faithfulness of the generated explanations. Figure~\ref{fig:pareto_fidelity} presents the trade-off between $Fidelity^{-}$ (y-axis) and $Fidelity^{+}$ (x-axis) over varying graph sparsity levels (5\%–95\%). An ideal explainer achieves high $Fidelity^{+}$ and low $Fidelity^{-}$, corresponding to points concentrated toward the lower-right region of the plot. As shown, the Top-2 Expert with LB scenario yields the most favorable distribution, followed closely by the Top-2 Expert without LB, indicating that these configurations produce the most reliable and self-contained subgraph explanations. Figure~\ref{fig:characterization_sparsity} reports the Characterization score across sparsity levels, which combines $Fidelity^{+}$ and $Fidelity^{-}$ into a unified measure of explanatory sufficiency and necessity. Higher values correspond to more faithful and comprehensive explanations. Consistent with the fidelity analysis, the Top-2 Expert with LB achieves the highest characterization values across all sparsity levels, with the Top-2 Expert without LB ranking second, confirming the stability and interpretive strength of these two routing strategies.

\section{Conclusion}\label{sec:conclusion}

This work introduced a routing-aware explainable MoE framework for malware detection using extracted CFGs. The proposed model integrates two levels of diversity: multi-statistic node encoding, which fuses neighborhood information through distinct pooling functions and degree reweighting, and expert-level specialization, where six experts capture complementary structural and behavioral perspectives of program graphs. The router selectively combines these experts through Top-$k$ routing with LB, achieving a balance between specialization and coverage. Empirical results demonstrated that the proposed Top-2 MoE with LB outperforms both single-expert GNN baselines and alternative MoE variants in terms of detection accuracy.
Beyond achieving competitive predictive performance, this study highlighted how incorporating routing dynamics into the explanation process leads to a more transparent interpretation of expert collaboration in MoE architectures. The fidelity-based and characterization-based analyses confirmed that routing-aware aggregation provides higher explanatory sufficiency and necessity compared to uniform or temperature-based routing. These findings suggest that explicitly modeling expert selection and LB not only benefits generalization but also strengthens the interpretability of graph-level decisions in malware analysis.

Future research will extend this work in several directions. First, we plan to generalize the framework to support multi-modal graph inputs, such as integrating control flow, API-call, and system-call graphs within a unified MoE architecture. Second, the router could be enhanced with adaptive temperature or uncertainty-driven gating mechanisms to further calibrate expert selection. Third, integrating local and global explainers or developing inherently interpretable experts could bridge post-hoc and intrinsic explainability. Finally, applying the proposed routing-aware interpretability approach to other graph-centric cybersecurity domains, including intrusion detection and vulnerability analysis, could validate its broader applicability.


\bibliography{sn-bibliography}

\end{document}